\documentclass[11pt,english,a4paper,twocolumn,twoside,floatfix]{article}
\usepackage{amstext,graphicx,geometry,color,parskip,babel,setspace,subfig,caption,floatrow,nccmath,bm}
\title{Functional Characteristics of Gene Expression Motifs with Single and Dual Strategies of Regulation}
\author{Mainak Pal$^{\dagger}$, Sayantari Ghosh\footnote{sghosh@kol.amity.edu} \:\:and Indrani Bose\footnote{indrani@jcbose.ac.in}\\
       \small $^{\dagger}$Department of Physics, Bose Institute\normalsize \\
       \small 93/1, Acharya Prafulla Chandra Road, Kolkata-700009, India\normalsize\\
       \small $^{*}$Amity Institute of Applied Sciences,\normalsize \\
       \small Amity University, Kolkata, India\normalsize }
       
\date{} 

\begin{document}
\maketitle
\normalsize 
\begin{abstract}
 Transcriptional regulation by transcription factors and post-transcriptional regulation by microRNAs constitute two major modes of regulation of gene expression. 
 While gene expression motifs incorporating solely transcriptional regulation are well investigated, the dynamics of motifs with dual strategies of regulation, i.e., 
 both transcriptional and post-transcriptional regulation, have not been studied as extensively. In this paper, we probe the dynamics of a four-gene motif with dual 
 strategies of regulation of gene expression. Some of the functional characteristics are compared with those of a two-gene motif, the genetic toggle, employing only 
 transcriptional regulation. Both the motifs define positive feedback loops with the potential for bistability and hysteresis. The four-gene motif, contrary to the 
 genetic toggle, is found to exhibit bistability even in the absence of cooperativity in the regulation of gene expression. The four-gene motif further exhibits a novel 
 dynamical feature in which two regions of monostability with linear threshold response are separated by a region of bistability with digital response. 
 Using the linear noise approximation (LNA), we further show that the coefficient of variation (a measure of noise), associated with the protein levels in the steady state, has 
 a lower magnitude in the case of the four-gene motif as compared to the case of the genetic toggle. We next compare transcriptional with post-transcriptional regulation 
 from an information theoretic perspective. We focus on two gene expression motifs, Motif 1 with transcriptional regulation and Motif 2 with post-transcriptional regulation.
 We show that amongst the two motifs, Motif 2 has a greater capacity for information transmission for an extended range of parameter values.
 \end{abstract}
 
 \section*{1. Introduction}
 \label{intro}
A living cell contains several thousands of genes the expression of a large fraction of which is 
regulated \cite{alberts1}. Two major strategies for the regulation of gene expression are transcriptional and 
post-transcriptional regulation, each with a distinctive mode of operation. In the case of transcriptional 
regulation, regulatory molecules bind the DNA to either activate or repress the initiation of transcription 
and thereby control the amount of proteins synthesized during translation. Post-transcriptional regulation is
 brought about by small non-coding RNAs (small RNAs/microRNAs) which bind the messenger RNAs (mRNAs) of the 
target gene  resulting in the degradation of the mRNAs and/or the inhibition of translation \cite{ambros2, levine0}. Small RNAs 
regulate bacterial gene expression whereas microRNAs are functional in eukaryotic cells.

The study of large-scale gene regulatory networks has so far been mostly confined to regulatory networks \cite{Alon, Alon1} which take into account only transcriptional regulation. The 
 connectivity structure of these networks reveals the existence of a number of gene expression motifs or substructures, e.g., positive and negative feedback loops, single input modules 
 and feed forward loops. In recent years, similar motifs have been identified in genome-scale regulatory networks with dual strategies of regulation, i.e., 
 the regulatory interactions include both transcriptional and post-transcriptional regulation \cite{Martinez, Shalgi, Tsang, Riba, Nitzan, Jolly}. Amongst the various types of motifs, the positive feedback loop has the 
 potential for the generation of multistability, i.e., the coexistence of more than one stable steady state for the same parameter values. Bistability, with two stable 
 steady states, provides the basis for cell-fate decisions between two alternative fates as in the case of cell differentiation in which a progenitor cell faces an either-or decision in choosing between two distinct cell lineages. 
 A biological switch utilizes bistability for the generation of the so-called ``OFF'' and ``ON'' states between which the switch operates. A well-known example of a biological switch is the genetic toggle, a two-gene motif, 
 in which the two genes mutually inhibit each other's expression through transcriptional regulation (figure 1({\it a})) \cite{Gardner}. The mutual inhibition (double negative feedback) defines a positive feedback loop and in 
 terms of the steady state concentrations [P$_1$] and [P$_2$] of the two proteins, the stable steady states correspond to [P$_1$]$\gg$[P$_2$] and [P$_2$]$\gg$[P$_1$] respectively. The two-gene motif appears extensively in several natural genetic networks including the ones governing cell-fate decisions \cite{Heinaniemi}.
 
 A number of recent studies \cite{Jolly, Lu, Tian, Bracken} have identified a pair of interconnected mutual-inhibition feedback circuits employing dual strategies of regulation. In a single feedback circuit, a microRNA and 
 a transcription factor (TF) mutually repress each other's synthesis. The microRNAs involved in the two feedback loops are {\it miR-34} and {\it miR-200} whereas the corresponding TFs are SNAIL and ZEB 
 respectively. The two interlinked feedback loops constitute the core of a gene regulatory network governing the epithelial to mesenchymal transition (EMT). The transition brings about 
 the conversion of epithelial to mesenchymal cells characterized by the loss of cell-cell adhesion and enhanced cell mobility. EMT plays crucial roles in the development of embryos and 
 the repair of tissues. Aberrantly regulated transitions result in cancer metastasis with primary tumor cells losing cell-cell adhesion and becoming migratory. Beyond two-gene motifs, there is limited knowledge 
 about more complex positive feedback loops with dual strategies of regulation. Figure 1({\it b}) provides a natural example of such a motif involving four genes \cite{Johnston, Bowerman}.  
 \captionsetup[figure]{labelformat=simple, labelsep=none}
\begin{figure}[h]
   \centering
     \includegraphics[scale=0.9,width=0.9\textwidth]{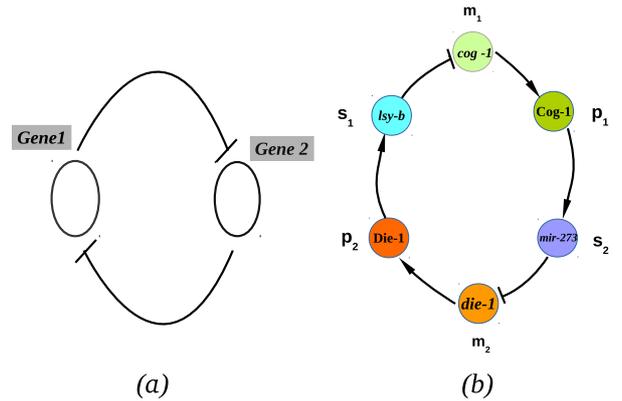}
    \caption{\small. Gene expression motifs: ({\it a}) Genetic toggle, a two-gene motif in which the genes mutually repress each other's expression.
     ({\it b}) Four-gene motif employing dual strategies of regulation, both transcriptional and post-transcriptional. The genes {\it die-1} and 
     {\it cog-1} synthesize the transcriptional regulators Die-1 and Cog-1 respectively. Die-1 (Cog-1) activates the expression of the {\it lsy-6}
     ({\it mir-273}) gene synthesizing the microRNA {\it lsy-6} ({\it mir-273}). The microRNA {\it lsy-6} ({\it mir-273}) targets the mRNA of the {\it cog-1}
      ({\it die-1}) gene. The meanings of the symbols {\it m}$_{i}$,  {\it s}$_{i}$,  {\it p}$_{i}$ \newline({\it i} = 1, 2) are explained in the text. Both the feedback loops in ({\it a}) and ({\it b}) define a positive feedback loop. The arrow signs indicate activation and the hammerhead symbols
      represents repression.
     \normalsize} 
     \end{figure}
The genes {\it die-1} and {\it cog-1} synthesize the  transcriptional regulators Die-1 and Cog-1 respectively. The genes {\it lsy-6} and {\it mir-273} synthesize microRNAs with 
Die-1 (Cog-1) activating the expression of the {\it lsy-6} ({\it mir-273}) gene. The microRNA {\it lsy-6} ({\it mir-273}) targets the mRNA of the {\it cog-1} ({\it die-1}) gene 
thereby inhibiting the synthesis of the Cog-1 (Die-1) protein. Thus the two transcriptional regulators Cog-1 and Die-1 mutually inhibit each other's synthesis in an 
indirect manner via post-transcriptional regulation by the microRNAs {\it lsy-6} and {\it mir-273}. The four-gene motif is part of the gene regulatory 
network governing the cell fate decision between two alternative neuronal fates, ASEL and ASER, in the nematode {\it C.elegans}. While the neuron pairs are left-right symmetric where 
position and morphology are concerned, they exhibit distinct asymmetries in the gene expression profile and neuronal function. The neurons
 ASEL and ASER are taste neurons which detect water-soluble sodium and chloride ions respectively during chemotaxis. The different neuronal fates enhance the 
ability of {\it C.elegans} to discriminate among environmental cues. A precursor neuron has the ASEL (ASER) fate when the Die-1 (Cog-1)
protein level is high and the Cog-1 (Die-1) level low. It should be noted that both the motifs, shown in figures 1(a) and 1(b), represent 
double-negative (mutual antagonism) feedback loops which effectively function as positive feedback loops. The dynamics of both the motifs allow for bistability, 
a prerequisite for the motifs to function as biological switches and also as regulators of cell fate decision.

The major objective of our paper is to investigate the functional characteristics of simple gene expression motifs with single and dual strategies for the 
regulation of gene expression. In the first part of our study (section 2), we undertake a comparative study of the gene expression motifs shown in 
figures 1(a) and 1(b). We consider both deterministic and stochastic dynamics of the motifs in order to identify the distinctive  operational features 
in the steady state. In the second part of our study (section 3), we consider two one-gene motifs designated as Motif 1 and Motif 2, In Motif 1, TFs repress the expression of the target gene 
whereas Motif 2 describes the post-transcriptional regulation of the target gene expression by  microRNAs. A regulatory module may be represented as an an input-output device with the regulatory molecules defining the input 
and the gene expression levels constituting the output \cite{Tkacik, Rhee, Tkacik1}. Due to  the stochastic nature of gene expression \cite{Kaern, Raj, Balazsi}, the input and output are not single levels but are best described as distributions around average values. The variability (noise) corrupts the 
fidelity of information transmission from the input to the output as the one-to-one correspondence between the input and the output is lost. The information-theoretic quantity mutual information (MI) provides knowledge of the amount of information that the value of one random variable yields about that of another.
\begin{table*}[htbp]
\caption*{ Table 1.\
Values of parameters in arbitrary units}
\begin{tabular}{c c c c c c}\hline\hline
  Parameter & figure 2 & figure 3(a)&figure 3(b) & figure 4 & figure 5\\ \hline \hline
  $\alpha{_{m_{1}}}$ & -& - & - & 15.00&-\\
  $\alpha{_{m_{2}}}$ & 15.00 & 30.00 & - & 15.00 & 0.950\\
  $\gamma{_{m}}$ & 0.200 & 0.200 & 0.200 & 0.200 & 0.800\\
  $\alpha{_{s}}$ & 1.000 & 1.000 & 1.000 & 1.000 & 1.000\\
  $\gamma{_{s}}$ & 0.080 & 0.080 & 0.080 & 0.080 & 0.080\\
  $\alpha{_{p}}$ & 0.100 & 0.100 & 0.100 & 0.100 & 1.400\\
  $\gamma{_{p}}$ & 0.080 & 0.080 & 0.080 & 0.080 & 0.080\\
  {\it t}$_{0}$  & 20.00 & 20.00 & 20.00 & 20.00 & 20.00\\
  {\it R} & 40.00 & 40.00 & 40.00 & 40.00&20.00\\
  $\mu{_{_1}}$& 0.200 & 0.200 & 0.200 & 0.200&0.200\\
  $\mu{_{_2}}$& 0.001 & 0.001 & 0.001 & 0.001&0.010\\
  {\it k}& 0.100 & 0.100 & 0.100 & 0.100&0.010\\
  $\tau$& 0.200 & 0.200 & 0.200 & 0.200&0.800\\
  
  \end{tabular}
  \end{table*}
The transmission of information through an input-output device can equivalently be represented as information flow through a channel. The 
channel capacity is defined to be the MI maximized over all possible input distributions and is a measure of the maximum amount of information 
that can be transmitted through the channel. In section 3 of the paper, we compute the channel capacity for the Motifs 1 and 2 to analyze transcriptional and 
post-transcriptional gene regulation from the viewpoint of information theory. Section 4 of the paper contains concluding remarks.
\captionsetup[figure]{labelformat=simple, labelsep=none}
 \begin{figure}[t]
\centering
  \includegraphics[scale=0.9,width=0.9\textwidth]{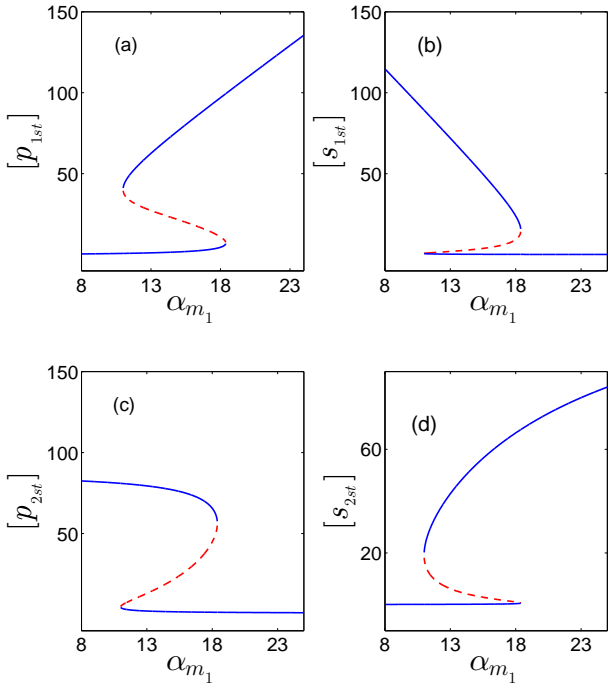}
  \caption{\small. Bistability in the steady states of the four-gene motif. The figures show the steady state concentrations of (a) [{\it p}$_{_1{{_{{st}}}}}$], 
  (b) [{\it s}$_{_1{{_{{st}}}}}$], (c) [{\it p}$_{_2{{_{{st}}}}}$],  (d) [{\it s}$_{_2{{_{{st}}}}}$] as a function of the mRNA synthesis rate constant 
  $\alpha_{_m{_{_1}}}$. The solid (dashed) lines represent stable (unstable) steady states. The parameter values are displayed in table 1.\normalsize}
 \end{figure}
 \section*{2. Dynamics of two-gene and four-gene motifs}
We first consider a four-gene motif with dual strategies of gene regulation, a specific example of which is shown in figure 1(b). In the motif, there are altogether 
four genes, two of which synthesize the mRNAs {\it m}$_1$ and {\it m}$_2$ and the other two synthesize the microRNAs {\it s}$_1$ and {\it s}$_2$. The mRNAs {\it m}$_1$ 
and {\it m}$_2$ are translated to yield the proteins {\it p}$_1$ and {\it p}$_2$. The microRNA {\it s}$_{i}$ binds the mRNA {\it m}$_{i}$ ({\it i} =1, 2) 
and targets it for degradation. The protein {\it p}$_1$ ({\it p}$_2$) activates the synthesis of the microRNA {\it s}$_2$ ({\it s}$_1$). The deterministic dynamics of the motif are described by the following set of differential equations: 
\begin{equation}
 \frac{ d[m{_{_1}}[}{dt}\ = \alpha{_{m_{_1}}}-\mu{_1}[m{_1}][s{_{_1}}] +\mu{_{_2}}[c{_{_1]}}-\gamma{_{m _{_1}}}[m{_{_1}}]
 \end{equation}
 
 \begin{equation}
 \begin{aligned}
  \frac{ d[s{_{_1}}]}{dt}\ = \alpha{_{s{_{_1}}}}+\frac{t{_{_1}}[p{_{_2}}]}{R{_{_1}}+[p{_{_2}}]}\ 
  -\mu{_{_1}}[m{_{_1}}][s{_{_1}}]+\mu{_{_2}}[c{_{_1}}]+ \\ k{_{_1}}[c{_{_1}}]-\gamma{_{s{_{1}}}}[s{_{_1}}]
 \end{aligned}
  \end{equation}
  
  \begin{equation}
  \frac{ d[c{_{_1}}]}{dt}\ = \mu{_{_1}}[m{_{_1}}][s{_{_1}}] - \mu{_{_2}}[c{_{_1}}] - k{_{_1}}[c{_{_1}}] - \tau{_{_1}}[c{_{_1}}]
  \end{equation}
 
 \begin{equation}
  \frac{ d[p{_{_1}}]}{dt}\ = \alpha{_{p_{_{1}}}}[m{_{_1}}] -\gamma{_{p_{_{1}}}} [p{_{_1}}]
  \end{equation}
  
  \begin{equation}
  \frac{ d[m{_{_2}}]}{dt}\ = \alpha{_{m_{_2}}} - \mu{_{_1}}[m{_{_2}}][s{_{_2}}] + \mu{_{_2}}[c{_{_2}}]-\gamma{_{m_{_2}}}[m{_{_2}}]
 \end{equation}
 
 \begin{equation}
 \begin{aligned}
  \frac{ d[s{_{_2}}]}{dt}\ =\alpha{_{s_{_2}}} +\frac{t{_{_2}}[p{_{_1}}]}{R{_{_2}}+[p{_{_1}}]}\ - 
  \mu{_{_1}}[m{_{_2}}][s{_{_2}}]+ \\ \mu{_{_2}}[c{_{_2}}]+k{_{_2}}[c{_{_2}}]-\gamma{_{s{_{2}}}} [s{_{_2}}]
  \end{aligned}
  \end{equation}
  
  \begin{equation}
  \frac{ d[c{_{_2}}]}{dt}\ = \mu{_{_1}}[m{_{_2}}][s{_{_2}}] - \mu{_{_2}}[c{_{_2}}] - k{_{_2}}[c{_{_2}}] - \tau{_{_2}}[c{_{_2}}]
  \end{equation}
 
 \begin{equation}
  \frac{ d[p{_{_2}}]}{dt}\ = \alpha{_{p_{_2}}}[m{_2}] -\gamma{_{p_{_2}}} [p{_{_2}}]
  \end{equation}
 The symbols {\it m}$_i$,  {\it s}$_i$,  {\it c}$_i$, {\it p}$_i$ ({\it i} = 1, 2) denote the molecular types as well as numbers and the 
 quantities within third brackets represent the concentrations. The modelling of the post-transcriptional regulation is based on 
 the mathematical models proposed in \cite{Mukherji, Schmiedel}. The models derive their strength from the fact that the theoretical results closely 
 match the experimental observations in single cell measurements.

The mRNAs are transcribed at the rates $\alpha{_{m_{_1}}}$ and $\alpha{_{m_{_2}}}$ and constitutively degraded at the rates $\gamma{_{m _{_1}}}$[{\it m}${_{_1}}$] 
 and $\gamma{_{m _{_2}}}$[{\it m}${_{_2}}$]. Proteins {\it p}${_{_i}}$ are translated from the free mRNA {\it m}$_i$ at the rate $\alpha{_{_p{_{_i}}}}$[{\it m}${_{i}}$] and degrade at the rate $\gamma{_{_p{_{_i}}}}[${\it p}${_{_i}}$] ({\it i} = 1, 2). The microRNA
 {\it s}$_i$ binds the target mRNA {\it m}$_i$ to form the mRNA-microRNA complex c$_{i}$ ({\it i} = 1, 2) which results in an accelerated degradation of the mRNA so that the translation of the mRNA into proteins is not possible. 
 The bound complex c$_{i}$ ({\it i} = 1, 2) forms with the rate constant $\mu_{1}$ and dissociates into free mRNA and free microRNA with the rate constant $\mu_{2}$. The mRNA {\it m}$_i$ in the bound complex c$_i$ degrades with the additional rate {\it k}$_i$ [c$_i$] ({\it i} = 1, 2) 
 while the microRNAs are recycled to the pool of free microRNAs after the mRNA degradation is completed. The bound complex C$_{i}$ has a natural degradation rate with rate constant $\tau_{i}$ ({\it i} = 1, 2). The free microRNA {\it s}$_i$ is synthesized at the rate $\alpha_{s_{i}}$ and 
 degraded at the rate $\gamma_{s_{i}}$ [{\it s}$_i$] ({\it i} = 1, 2). Protein {\it p}$_2$ ({\it p}$_1$) activates the synthesis of {\it s}$_1$ ({\it s}$_2$) with {\it t}$_1$ ({\it t}$_2$) being the maximum rate of synthesis and {\it R}$_1$ ({\it R}$_2$) being the corresponding Michaelis-Menten-type constant.
 The dynamics of the four-gene motif exhibit bistability in the steady state in an extended parameter regime. For the sake of simplicity, we consider the symmetric situation $\gamma_{m_{_1}}$=$\gamma_{m_{_2}}$=$\gamma_m$, 
 $\alpha_{s_{_1}}$=$\alpha_{s_{_2}}$=$\alpha_{s}$, $\gamma_{s_{_1}}$=$\gamma_{s_{_2}}$=$\gamma_{s}$, $\alpha_{p_{_1}}$=$\alpha_{p_{_2}}$=$\alpha_{p}$, $\gamma_{p_{_1}}$=$\gamma_{p_{_2}}$=$\gamma_{p}$, 
 {\it t}$_1$={\it t}$_2$={\it t}$_0$, {\it R}$_1$={\it R}$_2$={\it R}, $\mu_{1}$, $\mu_{2}$, {\it k}$_1$={\it k}$_2$={\it k} and $\tau_{1}$=$\tau_{2}$=$\tau$. 
 Figures 2({\it a})-2({\it d}) show the steady state concentrations [{\it p}${_{_{1st}}}$], [{\it s}${_{_{1st}}}$], [{\it p}${_{_{2st}}}$] and [{\it s}${_{_{2st}}}$] respectively as a function of the mRNA synthesis rate constant 
 $\alpha{_{_m{_{_1}}}}$. The parameter values used for the steady state solutions are displayed in table 1. The solid lines in figure 2 represent stable steady states 
 whereas the dashed lines correspond to branches of unstable steady states. The bistability in each case is accompanied by hysteresis, i.e., the transitions from the lower to the 
 upper stable steady state and from the upper to lower stable steady state occur at different values of the bifurcation parameter. The four-gene motif (figure 1), similar to the genetic toggle (figure 1({\it a})), defines a positive feedback loop in which the genes transcribing 
the mRNAs {\it m}$_1$ and {\it m}$_2$ mutually inhibit each other's expression. Consequently, the concentrations of the proteins {\it p}$_1$ and 
{\it p}$_2$ cannot be simultaneously high in a stable steady state. If the concentration of {\it p}$_1$ is high then that of {\it p}$_2$ is low 
and vice versa. A high concentration of {\it m}$_i$ (corresponding protein concentration is also high) is incompatible with a high concentration of 
the microRNA {\it s}$_i$ as {\it s}$_i$ targets {\it m}$_i$ for degradation ({\it i} = 1, 2) blocking protein synthesis in the process. The two stable steady states thus correspond to 
[{\it p}$_{1st}$], [{\it s}$_{2st}$] high, [{\it p}$_{2st}$], [{\it s}$_{1st}$] low and vice versa [figure 2]. The essential requirement 
 for the generation of bistability is the presence of positive feedback combined with an ultrasensitive response \cite{Chen}. The most common origin of ultrasensitivity 
 may be ascribed to cooperativity in the regulation of gene expression. A less explored source of ultrasensitivity lies in molecular sequestration \cite{Buchler, Buchler1} 
 in which the activity of a biomolecule A is compromised through sequestration due to the formation of an inactive complex with another biomolecule B. 
 In the case of the four-gene motif, the regulatory links constitute a positive feedback loop. The model exhibits bistability even in the abscence of cooperativity in gene expression. 
 The ultrasensitive response occurs through molecular sequestration with A representing the mRNA of the target gene and B representing the microRNA. In 
 the case of small RNA/microRNA-regulated gene expression, the ultrasensitivity arising from molecular sequestration is in the form of a linear threshold behaviour 
 \cite{Levine, Mukherji}. We consider the post-transcriptional regulation of the expression of a target gene by a microRNA. The target gene is not expressed if the mRNA synthesis rate 
 $\alpha{_m}$ is lower than a threshold value set by the microRNA synthesis rate $\alpha{_s}$. In this case, all the mRNA molecules form bound complexes 
 with the microRNAs. For $\alpha{_m}$ $ > $ $\alpha{_s}$, the number of mRNA molecules is larger than that of the microRNAs and some free mRNAs are available for translation 
 into proteins. The expressed protein level is linearly proportional to $\alpha{_m}$ $\textendash$  $\alpha{_s}$. The linear threshold behaviour has been demonstrated experimentally 
 in the cases of both small RNA and microRNA-regulated gene expression \cite{Levine, Mukherji}. Figure 3({\it a}) shows an example of linear threshold behaviour in the region of monostability of 
 the four-gene motif with the parameter values given in table 1. The linear threshold behaviour survives even in the presence of dual strategies of regulation of gene expression. The threshold location is now defined 
 by a relationship more complex than $\alpha{_m}$ $ = $ $\alpha{_s}$.  Figure 3({\it b}) exhibits how the steady state concentration of the protein {\it p}$_1$ changes in the enlarged parameter space $\alpha{_{m_{1}}}$-$\alpha{_{m_{2}}}$ from one region of monostability to another such region via a 
 region of bistability. The linear threshold behaviour indicates reversible response whereas the bistable response is digital (low-or-high) and hysteretic 
 in character. Figure 3({\it b}) captures the evolution of the steady state response of the four-gene motif from the linear threshold monostable to the digital bistable to finally the linear-threshold monostable response once again. This 
 type of steady state response has not been reported earlier in the case of microRNA-mediated regulation of gene expression. The parameter values for the figure are 
 given in table 1. An examination of figure 3({\it b}) shows that as the rate constant $\alpha_{m_{2}}$ decreases from a high value, the value of the upper bifurcation point (the value of $\alpha_{m_{1}}$ at which an abrupt transition 
 from the lower to the upper branch of stable steady states) also decreases. This is an expected result as the genes synthesizing the mRNAs {\it m}$_1$ and {\it m}$_2$ at the rates 
 $\alpha_{m_{1}}$ and $\alpha_{m_{2}}$ respectively, mutually inhibit each other's expression. Large values of $\alpha_{m_{2}}$ favour high (low) concentration of {\it m}$_2$ ({\it m}$_1$). Thus, as the value of $\alpha_{m_{2}}$ decreases, the transition to the 
 stable steady state with high {\it p}$_1$ concentration becomes more favourable, i.e., the upper bifurcation point shifts to lower values of $\alpha_{m_{1}}$. Steady state behaviour similar to that shown in figure 3({\it b}) is obtained if instead of the parameter space 
 $\alpha_{m_{1}}$-$\alpha_{m_{2}}$, one considers the parameter space $\alpha_{s_{1}}$-$\alpha_{s_{2}}$.

 In figure 4, we display the results of a parameter sensitivity analysis with respect to some of the parameters, $\alpha{_{m_{_1}}}$, $\gamma{_{m_{_1}}}$, $\alpha{_{p_{_1}}}$, $\gamma{_{p_{_1}}}$, 
 in the absence (figure 4({\it a})) and in the presence (figure 4({\it b})) of cooperativity in the transcriptional regulation of gene expression. The cooperativity is taken into account by 
 replacing the protein concentrations [{\it p}${_{_1}}$] and [{\it p}${_{_2}}$] in the second terms of equations (2) and (6) by [{\it p}${_{_1}}]^{n}$ and [{\it p}${_{_2}}]^{n}$ 
 respectively, where {\it n} is the Hill Coefficient. Similarly, the constants {\it R}${_{_1}}$ and {\it R}${_{_2}}$ are replaced by$R^{n}_{_1}$ and $R^{n}_{_2}$. 
 The set of parameter values shown in table 1 serves as the reference set. The value of each of the four parameters was varied separately in steps upto values that are 100 fold greater and lower than the reference value listed in table 1. 
 Figure 4 shows the log-fold change in parameter value within the range -2 to +2 for a specific parameter (marked on the {\it x} axis) and identifies the region of bistability, shaded black, in the parameter range. As figure 4(a) 
 clearly demonstrates, bistability occurs even in the abscence of cooperativity. With cooperativity taken into account ({\it n} = 2), the region of bistability becomes more extensive 
 (figure 4(b)). In this case, both cooperativity and molecular sequestration contribute to the generation of the ultrasensitive response. In the presence of cooperativity, the steady state response becomes sharper in the region of monostability. 
 \captionsetup[figure]{labelformat=simple, labelsep=none}
 \begin{figure*}[t]
  \centering
   \centering
 \subfloat[]{%
 \includegraphics[height=6cm,width=6cm]{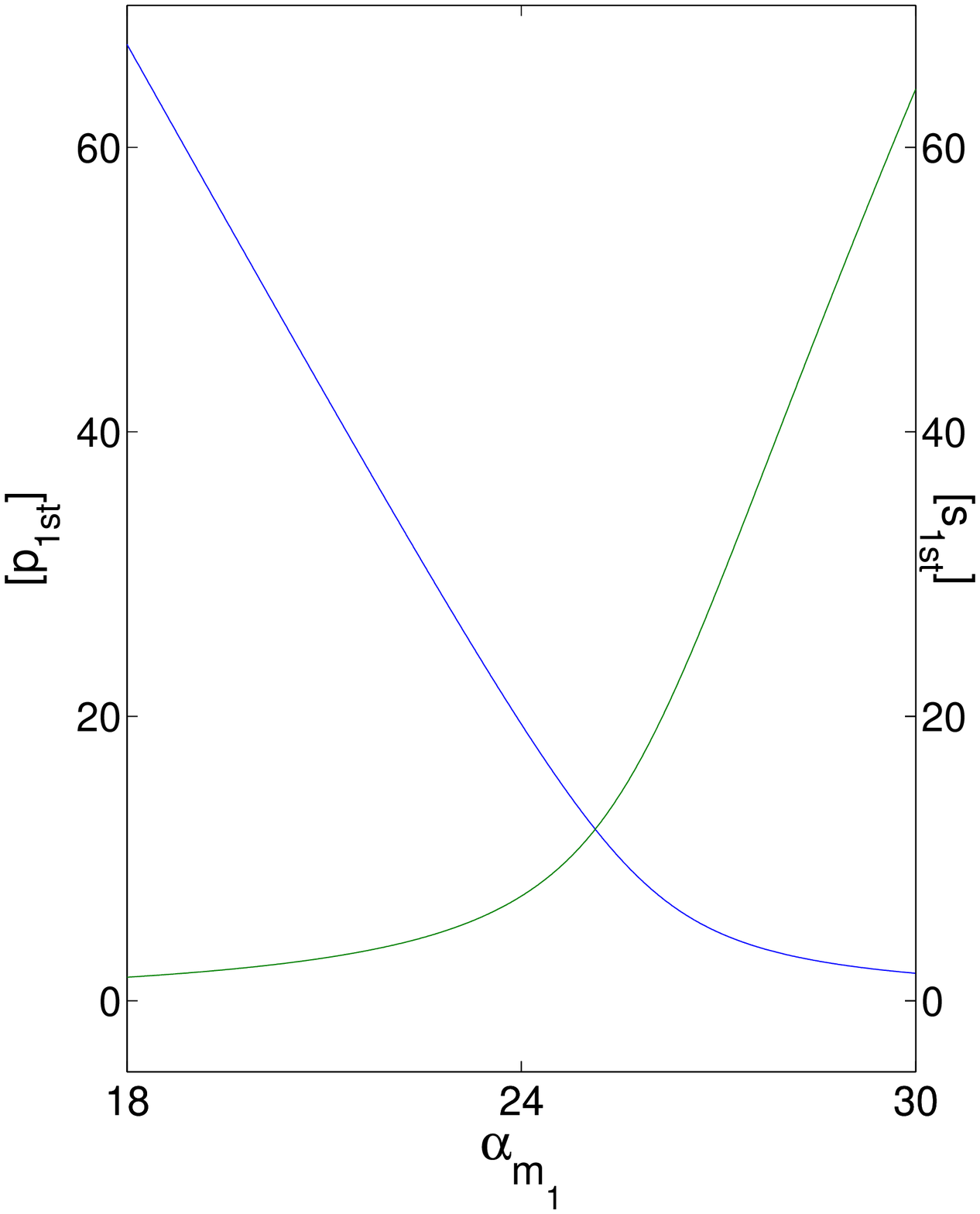}}
 \quad
 \subfloat[]{%
 \includegraphics[height=8cm,width=11.5cm]{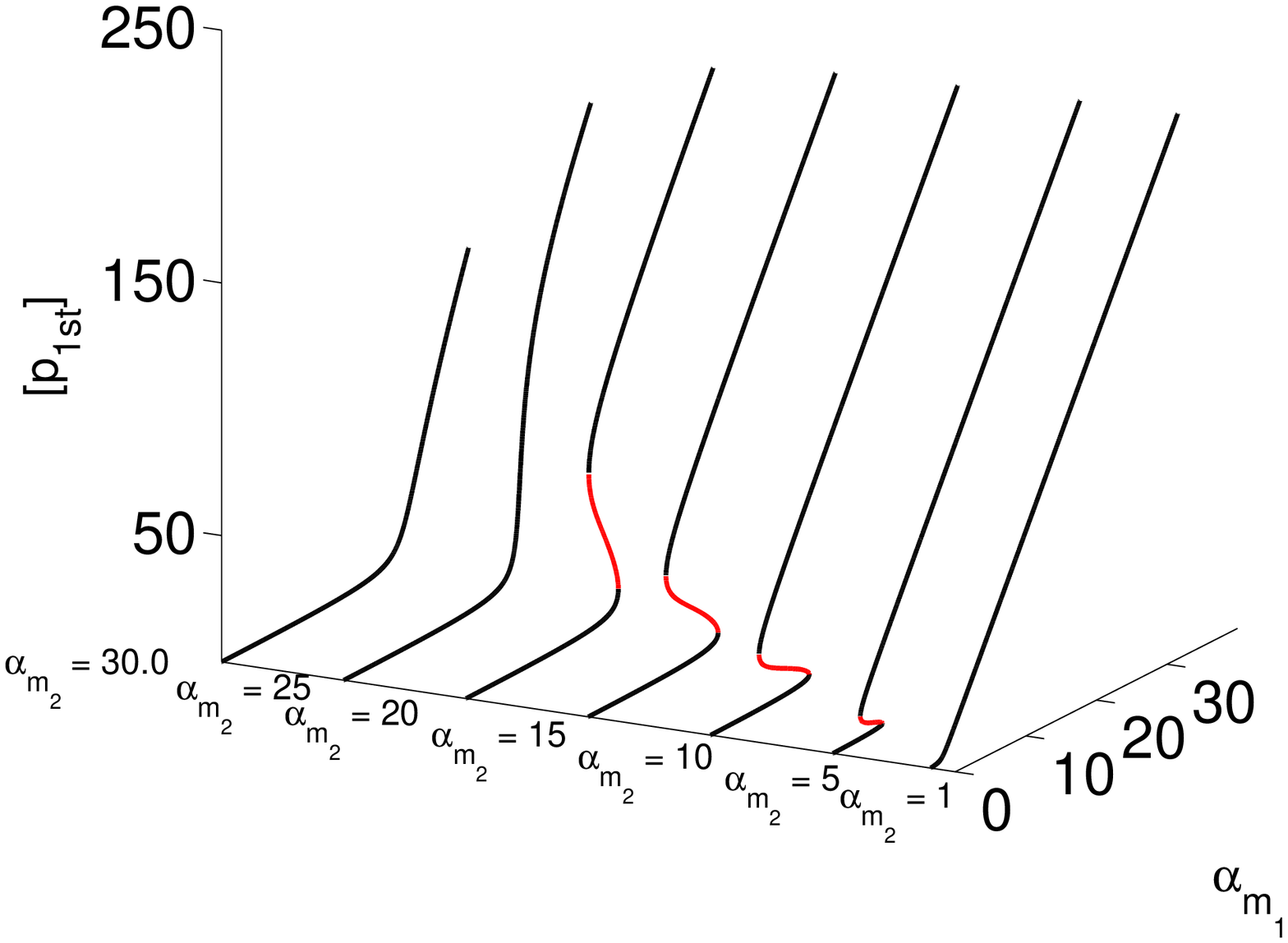}}
   \caption{\small. (a) Linear threshold behaviour in the region of monostability of the four-gene motif. (b) Variation of the steady state protein concentration in the enlarged parameter space 
     $\alpha _{m_{1}}$-$\alpha _{m_{2}}$ from one region monostability to another such region via a region of bistability. The diagram shows how linear threshold behaviour 
     evolves into bistability accompanied by hysteresis. The parameter values are displayed in table 1.
     \normalsize}
     \end{figure*}  
 
 One important advantage of microRNA-mediated regulation of gene expression is that microRNAs (as well as small RNAs) are known to suppress fluctuations in the target gene's protein levels \cite{Levine, Schmiedel, Ostella, Siciliano}. Cell fate decisions 
 are, in general, controlled by bistable genetic circuits \cite{Balazsi}. The two stable steady states are defined in terms of the steady state concentrations of key regulatory 
 proteins. The stable steady states correspond to the two stable phenotypes that a cell has to choose between. Fluctuations in the protein levels can bring about 
 undesirable transitions between the alternative states resulting in a loss in the stability of the phenotypes. A robust biological switch operating 
 between two stable steady states eliminates to a large extent the random transitions between the states. A major outcome of stochastic gene expression is that 
 protein production occurs in random bursts rather than in a continuous manner \cite{Kaern, Raj, Yu}. A single mRNA has a finite lifetime during which the mRNA is repeatedly translated 
 to yield a burst of proteins. The average size {\it b} of a protein burst is given by {\it b} = $\frac{\alpha{_{_p}}}{\gamma_{_m}}$ where $\alpha{_{_p}}$ is the translation rate and 
 $\frac{1}{\gamma{_{_m}}}$ is the average lifetime of the mRNA. In the case of post-transcriptional regulation, the microRNAs promote mRNA degradation ($\gamma{_{_m}}$  is increased) and/or suppress translation ($\alpha_{_p}$ decreases). 
 The outcome is a reduction in the protein burst size, i.e., diminished protein fluctuations. In a two-step model of gene expression, the noise at the protein level is quantified in terms of the coefficient of variation (CV$_{_p}$) which is the ratio of the 
 standard deviation and mean $\langle${\it p}$\rangle$ \cite{Thattai}. The expression for C$V_{_p}$ in the steady state is C$V_{_p}^{2}$ = $\frac{1+b}{\langle{\it p}\rangle}$. A reduction in the average burst size lowers the amount of relative 
 fluctuations in the protein level. We compute the CV$_{_p}$  for both the two-gene and four-gene motifs, shown in figure 1, to ascertain which motif is less noisy. We use the linear noise approximation (LNA) to 
 the Master equation \cite{Van, Elf, Pal} governing the stochastic dynamics of the motifs. In the following, we outline the major steps in the computational procedure. We consider 
{\it N} distinct chemical species which participate in {\it R} chemical reactions. The concentrations of the chemical species are given by the variables {\it x}$_i$, {\it i} = 1, $\cdots$ {\it N}. 
The state of the dynamical system is represented by the vector \textbf{x} = ({\it x}$_1$, {\it x}$_2$, $\cdots$, {\it x}$_N$)$^{T}$ where {\it T} 
denotes the transpose. The state changes as a function of time due to the occurrence of the chemical reactions. Let \textbf{S} be the stoichiometric matrix with elements 
{\it S}$_{ij}$, {\it i} = 1, 2, $\cdots$, {\it N}, {\it j} = 1, 2, $\cdots$, {\it R}. The interpretation of {\it S}$_{ij}$ is that the number of molecules of the chemical species {\it i} 
changes from {\it X}$_i$ to {\it X}$_i$ + {\it S}$_{ij}$ when the {\it j} th reaction occurs. One also defines a reaction propensity vector \textbf{f}(\textbf{x}) = 
({\it f}$_1$(\textbf{x}), $\cdots$, {\it f}$_R$(\textbf{x}) )$^T$. The deterministic dynamics of the system are generated by the rate equations 
\begin{figure}[h]
   \centering
     \includegraphics[scale=1.0,width=1.0\textwidth]{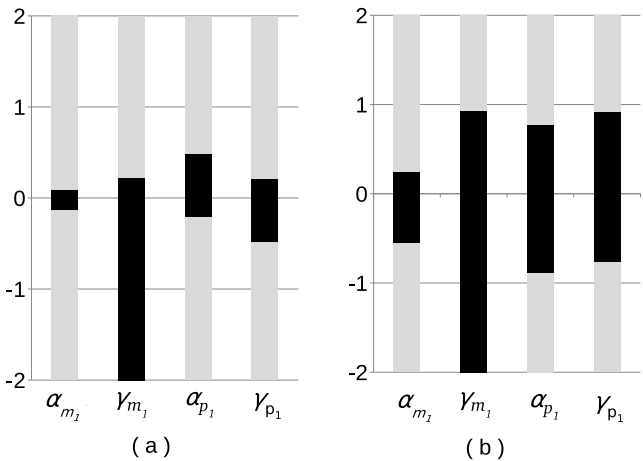}
    \caption{\small. Parameter sensitivity analysis of four-gene motif in the (a) absence and (b) presence of cooperativity in the 
    transcriptional regulation of gene expression. The value of each of the four parameters tested has a 100 - fold variation greater 
    and lesser than the reference value shown in the table 1. The {\it x} - axis shows the parameter varied and the {\it y} - axis depicts the 
    log - fold variation. The region of bistability is shaded dark.
     \normalsize} 
     \end{figure}
\begin{equation}
 \frac{dx_{i}}{dt} = \sum_{j=1}^{R} S_{ij} f_{j} (x) \:\:\:\:\:\:\: ( i = 1, 2, \cdots, N)
\end{equation}

The steady state vector \textbf{x$_s$} is determined from the condition $\dot{\textbf{x}}$ = 0, i.e., \textbf{f}(\textbf{x$_s$}) = 0. 
 
In the steady state and under the LNA, the covariance of the fluctuations about the deterministic steady state is given by the fluctuation-dissipation (FD) relation 
\begin{equation}
 \textbf{J}\textbf{C} + (\textbf{J}\textbf{C})^T +\textbf{ D} = 0
\end{equation}
 where J is the Jacobian matrix, \textbf{C}  = $\langle$ $\delta$\textbf{x} $\delta$\textbf{x}$^T$ $\rangle$ is the covariance matrix, 
 the diagonal elements of which are the variances, and \textbf{D} is the diffusion matrix. The matrix 
 \textbf{D} is given by \textbf{D} = \textbf{S}\:\: diag (\textbf{f}(\textbf{x}))\:\: \textbf{S}$^T$
\begin{figure}[t]
   \centering
     \includegraphics[scale=0.6,width=0.6\textwidth]{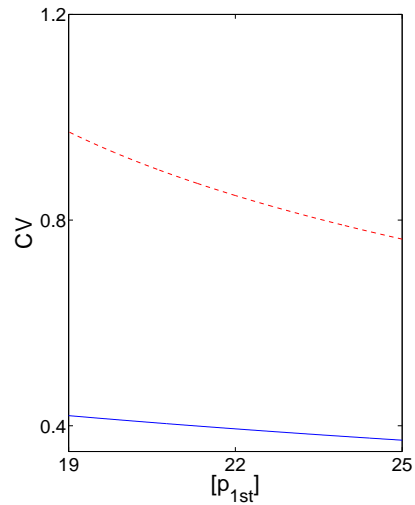}
    \caption{\small. Comparison of the coefficient of variation (CV) versus the steady state protein concentration [{\it p}$_{1}$] in the cases of the genetic 
    toggle (red dashed line) and the four-gene motif (blue line) .
     \normalsize} 
     \end{figure}
where diag(\textbf{f}(\textbf{x})) is represented by a diagonal matrix with the elements {\it f}$_{j}$ (\textbf{x}), {\it j} = 1, 2,  $\cdots$, {\it R}. Once the matrices 
\textbf{J} and \textbf{D} are specified, the elements of the covariance matrix, C$_{ij}$ = $\langle$ $\delta${\it x}$_i$ $\delta${\it x}$_j$ $\rangle$  ({\it i}, {\it j} = 
1, 2,  $\cdots$, {\it N}), with C$_{ii}$ ({\it i} = 1, 2, $\cdots$, {\it N}) denoting the variance, can be calculated using the FD relation (10). Under the LNA, the 
mean value of {\it x}$_i$ is given by the deterministic steady state value so that the CV associated with {\it x}$_i$ can be computed. In the case of the two-gene (four-gene) 
motif {\it N} = 4 ({\it N} = 8). In the Appendix, the differential rate equations governing the deterministic dynamics of the two-gene motif, the genetic toggle, 
are displayed in equations (A.1)-(A.4). The set of reaction schemes for both the two-gene and four-gene motifs and the matrices \textbf{S} and \textbf{f}(\textbf{x}) are also shown. Figure 5 exhibits the plots of the CV versus the protein 
mean value in the steady state in the cases of both the genetic toggle and the four-gene motif. The first motif employs only transcriptional regulation whereas the second motif 
incorporates the dual strategies of both transcriptional and post-transcriptional regulation of gene expression. The motif with dual strategies of regulation is clearly less noisy than 
the motif with only transcriptional regulation.The parameter values for the four-gene motif are given in table1. The parameters for the genetic toggle are 
{\it j}${_{M{_{_2}}}}$ = 0.9, $\delta_{M{_{_1}}}$ = 0.4, $\delta_{M{_{_2}}}$=0.4, {\it j}${_{p{_{_1}}}}$ = 1.0, {\it j}${_{p{_{_2}}}}$ = 1.0, $\delta_{p{_{_1}}}$ = 0.6, $\delta_{p{_{_2}}}$ = 0.6, $\beta_{1}$ = 70.0, 
$\beta_{2}$ = 70.0, K$_{1}$ = 0.01, K$_{2}$ = 0.01 ({\it j}${_{_{M{_{_1}}}}}$ is the parameter which is varied). In stochastic gene expression, the frequency {\it a} of the transcriptional protein bursts is given by 
{\it a} = $\frac{\alpha_{m}}{\gamma_{p}}$ where $\alpha_{m}$ is the transcription rate and $\gamma_{p}$ the protein degradation rate constant \cite{Friedman}. As discussed in \cite{Ostella}, 
transcriptional regulation modifies $\alpha_{m}$ thereby changing the burst frequency {\it a}. The steady state mean protein level $\langle${\it p}$\rangle$ = {\it b}{\it a}. 
Since C$V_p^2$ = $\frac{1+{\it b}}{\langle p \rangle}$, for the same degree of repression brought about transcriptionally or post-transcriptionally, the microRNAs are more effective 
in reducing the CV$_p$ as they diminish the burst size {\it b}.

\section*{3. Regulation strategies: Information theory perspective}

In this section, we compute the channel capacities of the input-output devices. representing Motifs 1 and 2 respectively.  The 
MI between the regulatory molecule distribution {\it p}({\it x}) and the output protein distributions {\it p}({\it y}) is given by
\begin{equation}
 I (x\:; y) = \iint \:\:dx dy \:\:p(x, y) \:\:log_{_2} \frac{p(y|x)}{p(y)}
 \end{equation}
where the variable {\it x} represents the concentration of the regulatory molecules and the variable {\it y} denotes the concentration of 
the proteins synthesized by the target gene, {\it p}({\it y}$\mid${\it x}) is the conditional probability distribution and the joint 
distribution {\it p}({\it x},{\it y})={\it p}({\it y}$\mid${\it x}) {\it p}({\it x}). The MI is measured in bits and has a clear physical 
interpretation. For the value of MI = I bits, 2$^{I}$ levels of {\it y} can be distinguished, in the presence of noise 
in the channel, given the value of the input {\it x}. Equivalently since the MI is a symmetric function of {\it x} and {\it y}, 2$^{I}$ input levels can be distinguished given the value of the output {\it y}. In the small noise approximation \cite{Tkacik, Tkacik1, Bialek}, the channel capacity for information transmission 
is given by the MI maximized over all possible distributions, {\it p}({\it y}), of the output signal. The optimal distribution has the form 
\begin{equation}
 p{_{opt}} = \frac{1}{Z}\: \frac{1}{\sigma_{y}}
\end{equation}
where $\sigma_{y}$ (the standard deviation) is a measure of noise in the output and Z normalises the distribution ,i.e.,
\begin{equation}
  Z = \int \frac{dy}{\sigma_{y}}
 \end{equation}
With the result (equation (12)) for the optimal distribution, the optimal MI or the channel capacity is given by 
\begin{equation}
 I{_{opt}} = log{_{_2}}\:[\: \frac{Z}{\sqrt{2 \pi e}}]
\end{equation}
We now compute the channel capacities for Motif 1 and Motif 2. In the case of Motif 1, the expression for $\sigma_{y}$ is known \cite{Tkacik, Tkacik1, Bialek} and we simply quote the result. In the case of Motif 2, we utilize the expression for $\sigma{_{y}}$, as derived in Ref. \cite{Schmiedel}, in the case of the post-transcriptional regulation of the expression of a 
single gene. The total noise in the steady state as measured by the CV, has two components, intrinsic and extrinsic: CV$_{tot}^{2}$ = CV$_{int}^{2}$ + CV$_{ext}^{2}$
with CV$_{int}$ being $\frac{\sigma_{y}}{mean}$. The mathematical modelling in \cite{Schmiedel} made the prediction that microRNA-mediated regulation affects the intrinsic and extrinsic noise in different ways. The 
microRNA-regulated gene expression is characterized by reduced intrinsic noise in comparison to the case of an unregulated gene at the same protein expression levels. 
The reduction in intrinsic noise,  $\frac{\eta_{int}}{\eta_{int}^{reg}}$,  is approximately equal to $\sqrt{r}$ where {\it r} denotes the microRNA-mediated fold repression 
( {\it r} = $\frac{[p_{st}^{unreg}]}{[p_{st}^{reg}]}$), where {\it p}$_{st}$ denotes the steady state protein concentration. The origin of extrinsic noise lies in the fluctuations in the pool of 
microRNAs. As shown in \cite{Schmiedel}, the combined effects of reduced intrinsic and additive extrinsic noise results in decreased total noise when expression levels are low and increased total noise at 
high expression levels. The prediction of the mathematical modelling was confirmed in single cell experiments \cite{Schmiedel}. In our study, we consider only the intrinsic component of the total noise at the output as for low and intermediate expression 
levels, the intrinsic component is dominant. In the small noise approximation, one requires a knowledge of only the variance $\sigma_{y}^{2}$ of the output distribution to compute the optimal output distribution and 
the channel capacity. We follow the notation of section 2 for the different parameters and quantities associated with microRNA-regulated gene expression. For Motif 2, the input variables {\it x} and {\it y} are [{\it s}], 
the concentration of free microRNAs, and [{\it p}], the concentration of target gene proteins respectively. From equation (20) of the Supplementary Materials of Ref. \cite{Schmiedel}, the standard deviation of the output distribution is given by 
\begin{equation}
 \sigma_{p} (\overline{p}) = \sqrt{[p{_{0}}]\overline{p} \: + \frac{b_{0}\: [p{_{0}}] \:\overline{p}\:(1-R_{s})}{(1-R_{s}\: F)^{2}}}
\end{equation}
where $\overline{p}$ is the normalized protein concentration, $\overline{p}$ = $\frac{[p]}{[p_{0}]}$ ( [{\it p}$_{0}$] is the maximum value of 
[{\it p}] corresponding to the unrepressed condition ) which has values in the range [0, 1]. The quantity {\it b}$_{0}$ = $\alpha_{p}$/$\gamma_{m}$ 
( in the notation of section 2 and with the suffixes omitted) is the average number of proteins synthesized from a mRNA during its 
lifetime, in the absence of any regulation. The quantity {\it R}$_s$ = 1-$\frac{[p]}{[p_{0}]}$ = 1-$\overline{p}$ denotes the repression strength of the microRNA-mediated regulation with 
0 $\leq${\it R}$_{s}$ $\leq$ 1. The values {\it R}$_s$=0 and 1 indicate no repression and full repression respectively of protein synthesis. 
\begin{figure*}[t]
   \centering
   \subfloat[]{%
     \includegraphics[scale=0.42,width=0.42\textwidth]{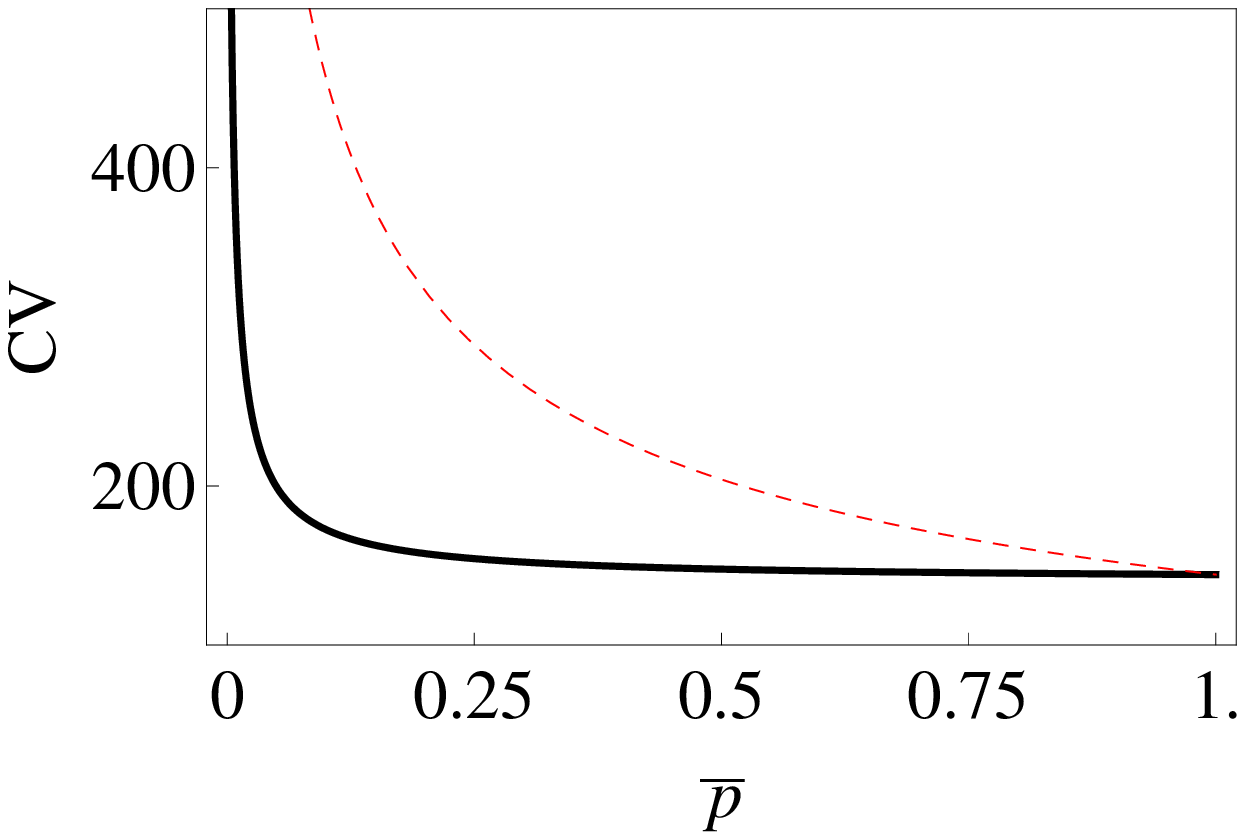}}
   \quad
   \subfloat[]{%
     \includegraphics[scale=0.42,width=0.42\textwidth]{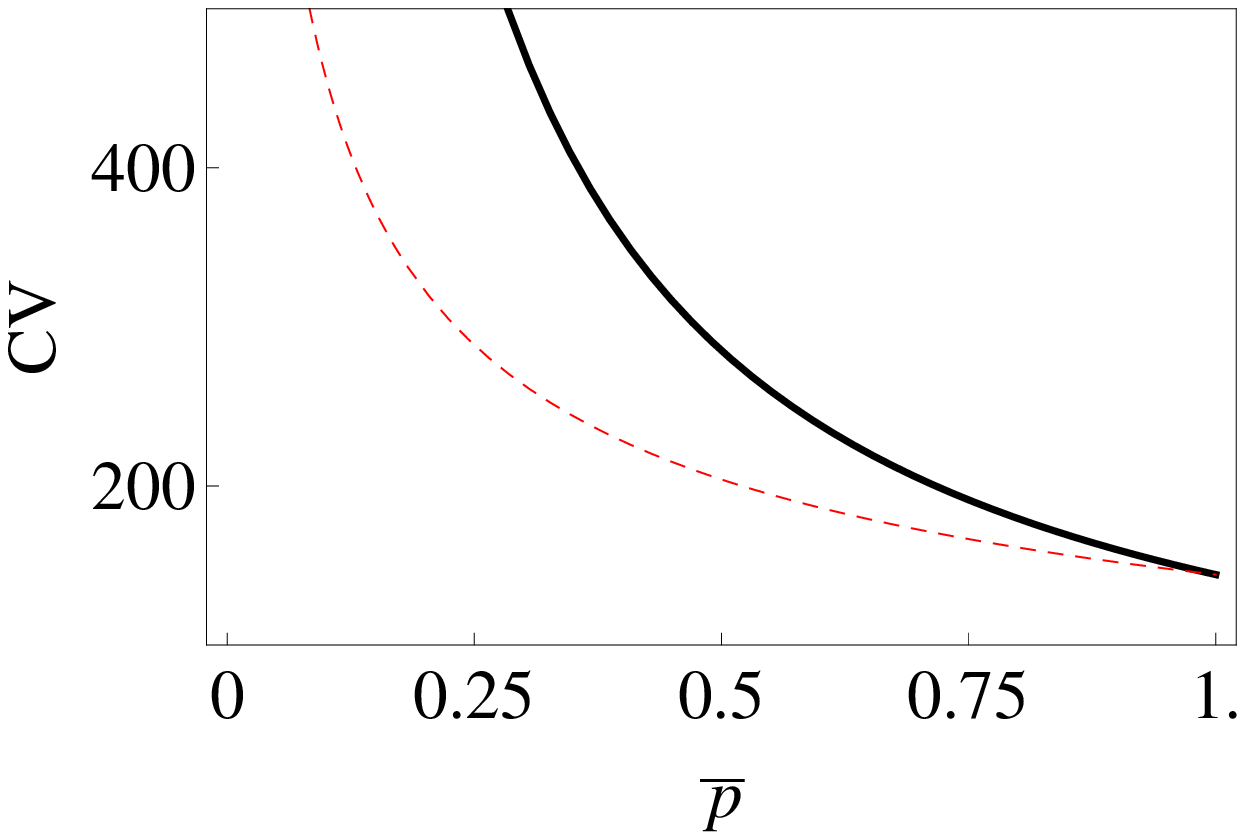}}
     \caption {\small.\:Coefficient of variation (CV) versus $\overline{p}$ for Motif 1 (dashed line) and Motif 2 (solid line) for ({\it a}) F$\simeq$ 0 and \newline({\it b}) F$\simeq$ 1\normalsize }
    \end{figure*}
The quantity F is given by
 \begin{equation}
 F=1-\frac{[s]}{[s_{tot}]}
 \end{equation}
 where [s] is the concentration of free microRNAs and [s$_{tot}$] is the total concentration of microRNAs, a fraction of which is free and the rest form bound complexes with the mRNAs. From (13), Z is given by the integral 
\begin{equation}
Z=[p_{0}]\int_{0}^{1} \frac{d\overline{p}}{\sqrt{[p_{0}]\:\overline{p}\:+\frac{b_{0}\:[p_{0}]\:\overline{p}\:(1-R_{s})}{(1-R_{s} F)^{2}}}} 
\end{equation}
The value of the protein concentration [p$_{0}$]=$\frac{\alpha_{p}\:\alpha_{m}}{\gamma_{p}\:\gamma_{m}}$. Figures 6(a) and 6(b) show the plots (solid lines) of 
the CV, $\frac{\sigma_{p}(\overline{p})}{\overline{p}}$, versus $\overline{p}$ for F$\simeq$ 0 and F$\simeq$ 1 respectively. The case F$\simeq$ 0 corresponds to the experimental situation in Ref. \cite{Schmiedel} with most of the microRNAs being free. 
For F$\simeq$1, almost all the microRNAs form bound complexes with the mRNAs. The parameter values for the plots are 
$\alpha_{m}$= 0.01, $\gamma_{m}$= 0.0058, $\alpha_{p}$= 0.1155 and $\gamma_{p}$= 0.0002 (the parameter values are the same as reported in Ref. \cite{Thattai}). From (14) and (17), the channel capacity I$_{opt}$= 2.91 (F$\simeq$ 0) and 
I$_{opt}$= 0.74 (F$\simeq$ 1). For (F$\simeq$ 0), the term R$_{s}$F in equation (17) may be neglected, i.e., the factor (1-R$_{s}$F) is replaced by 1. Similarly, for F$\simeq$ 1, 1-R$_{s}$F is replaced by 1-R$_{s}$.  In the case of Motif 1, $\sigma_{p}^{TF}(\overline{p})$ is given by \cite{Tkacik, Thattai} 

\begin{equation}
 \sigma_{p}^{TF}(\overline{p})=\sqrt{\overline{p}\:[p_{0}]\:(1+b_{0})}
\end{equation}

Figures 6({\it a}) and ({\it b}) show the plots (dotted lines) of the CV versus $\overline{p}$ for F$\simeq$ 0 and F$\simeq$ 1 respectively. 
In the first case, the CV (Motif 2) is less than the CV (Motif 1) and the situation is reversed for F$\simeq$ 1. The channel capacity for Motif 1 is given by 
I$_{opt}^{TF}$= 1.74. When F$\simeq$ 0 (F$\simeq$ 1), I$_{opt}$ is greater (less) than 
\begin{figure}[h]
 \centering
 \includegraphics[scale=0.8,width=0.8\textwidth]{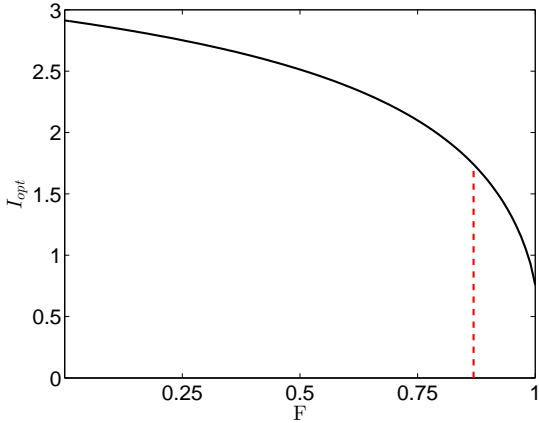}
 \caption {\small.\:\:Plot of I$_{opt}$ versus F. The dotted line corresponds to F=F$_{c}$=0.8961 above which I$_{opt}^{TF}$=1.74 is grater than I$_{opt}$.\normalsize }
\end{figure}
I$_{opt}^{TF}$. Figure 7 shows a plot of I$_{opt}$ versus F for Motif 2 with $\alpha_{p}$, $\gamma_{p}$, $\alpha_{m}$, $\gamma_{m}$  having the same values as in the case of figure 6. The channel capacity is found to decrease as F 
increases. The dotted line corresponds to the value of F=F$_{c}$=0.8961 above which I$_{opt}^{TF}$ is greater than I$_{opt}$ with I$_{opt}$=I$_{opt}^{TF}$ at 
F=F$_{c}$. Thus, over a considerable range of F values, Motif 2 has a greater channel capacity than Motif 1. The difference is maximal when F=0.
\section*{4. Summary and discussion}
Regulatory networks governing gene expression dynamics are characterised by the appearance of different types of motifs. The motifs emerged through evolutionary processes 
because of their functional superiority over more random substructures. In this paper, we study a four-gene motif with dual strategies of regulation to identify the 
distinctive functional features which enhance its performance as a biological switch and/or its role in cellular decision making in comparison with the genetic toggle, a two-gene motif 
employing transcriptional regulation. The four-gene motif is a component of the gene regulatory network controlling the cell fate decision between two alternative neuronal fates in 
{\it C.elegans}. The motif is found to be bistable over a wide parameter regime (figure 4). The occurence of bistability, even in the absence of cooperativity in regulation, 
is due to the combined effects of positive feedback and ultrasensitivity generated by molecular sequestration \cite{Chen, Buchler, Buchler1}. The genetic toggle, on the other hand requires cooperativity in 
regulation to achieve bistability. Post-transcriptional regulation, operative in the four-gene motif, involves molecular sequestration in the form of the bound microRNA-mRNA 
complex which is responsible for the generation of ultrasensitivity in the steady state concentration of mRNAs/proteins as a function of the mRNA synthesis rate constant. We provide another example of 
molecular-sequestration based bistability without cooperativity. In {\it Mycobacterium tuberculosis}, the stress response network consists of multiple positive feedback loops involving the two-component system MprAB and the gene {\it sigE} 
synthesizing the alternative sigma factor $\sigma ^{E}$. The positive feedback loops by themselves cannot generate bistability in biochemically realistic parameter ranges. Bistability is obtained on inclusion of the effect of molecular sequestration 
of $\sigma ^{E}$ by its anti-$\sigma$ factor RseA \cite{Tiwari}. We now briefly discuss some other examples of bistability without cooperativity. Bacterial two component systems regulate global responses to various types of stress. A two-component system consists of a sensor kinase (SK) 
and a response regulator (RR). The phosphorylated SK transfers the phosphate group to the RR which mediates the response to the stress signal. In general, the SK has a bifunctional character, it can act as a kinase in the phosphorylated state and as a phosphatase (dephosphorylating the RR) 
in the unphosphorylated state. The two-component system can exhibit bistability if the unphosphorylated SK and RR form a dead-end complex \cite{Igoshin} and (or) the phosphatase activity of the SK is much reduced (abolished) \cite{Ram}. A synthetic circuit constructed in {\it E.coli} with autoregulating 
T7 RNA polymerase involves a non-cooperative positive feedback which by itself cannot generate bistability. The expression of T7 RNA slows down cell growth thus reducing protein dilution and creating an implicit positive feedback loop. The two loops, in combination, can give rise to bistability \cite{Tan}.
Several toxin-antitoxin operons are regulated according to the toxin/antitoxin ratio through a mechanism termed `` conditional cooperativity''. The conditional regulation can give rise to bistability for a wide range of parameter values \cite{Cataudella}. Additionally, stochastic effects can generate bistability/bimodal gene expression 
without the requirement of cooperativity in the regulation of gene expression \cite{Lipshtat, Ochab}. As discussed in section 2, cooperativity in the regulation of gene expression has the effect of extending the region of bistability, a feature pointed out in several earlier studies \cite{Tiwari1, Huang, Chickarmane}.  

The ultrasensitivity in the case of post-transcriptional regulation is in the form of a linear threshold behaviour \cite{Levine, Mukherji}. The four-gene motif, incorporating both transcriptional and post-transcriptional regulation also exhibits linear threshold behaviour (figure 3({\it a})). 
The theoretical prediction could be tested in the actual experimental setting of Ref.\cite{Johnston} or by constructing a synthetic genetic circuit, as in the case of the genetic toggle \cite{Gardner}. The linear 
threshold behaviour, by setting a threshold for activated response, prevents spurious switchings to the active state. In the latter state, the response is analog, i.e., graded. 
The threshold is further tunable through the modifications of the mRNA and microRNA synthesis rate constants. The four-gene motif, because of its enlarged parameter space, exhibits a 
novel steady state response (figure 3(b)) which has not been reported earlier. The region of bistability separates two regions of monostability. In this region, 
response is digital because of the discontinious changes in response at the bifurcation points. The response further exhibits hysteresis whereas the response in the monostable regions is 
reversible. The four-gene motif thus combines the advantages of the linear threshold behaviour with bistability in order to mediate cellular decision making or act as a biological 
switch. The linear threshold behaviour provides the basis for an ultrasensitive switch with reversible response. The genetic toggle exhibits bistability 
in an extended parameter region with two monostable regions separating the region bistability. In the monostable regions, the response is graded and featureless. The motifs studied in Refs. \cite{Jolly, Lu, Tian, Bracken} exhibit tristability and thus can function as a three-way switch in contrast to the two-way switch behaviour of the motifs studied in our paper. 
The two-gene and four-gene motifs studied in the paper exhibit monostability or bistability depending on the parameter regime. Motifs which include negative feedback loops and/or introduce time delays in gene expression can exhibit oscillations in protein concentrations \cite{Tiana, Tiana1}. Small RNAs have been shown to establish delays and temporal thresholds in gene expression \cite{Legewie} as 
well as generate oscillations in protein concentrations \cite{Liu}. Motifs governed by transcriptional regulation are also known to exhibit oscillations \cite{Tiana, Tiana1}. It will be of interest to compare the functional characteristics of motifs incorporating transcriptional and post-transcriptional modes of gene regulation respectively, in generating oscillatory dynamics. 

Post-transcriptional regulation mediated by microRNAs has the desirable feature of reducing the variability or noise in the target gene protein levels \cite{Levine}. The reduction is 
mainly due to the diminished burst sizes in protein production. Reduction in protein noise, as compared to the case of unregulated gene expression, has been demonstrated experimentally 
in the cases of simple motifs with dual strategies of regulation of gene expression \cite{Ostella, Siciliano}. A more recent study \cite{Schmiedel} combines mathematical modeling with single cell experiments to establish that 
microRNA-regulated gene expression results in lower intrinsic noise as compared to the case of unregulated gene expression. We compute the CV of the protein levels of a target gene in the steady state of the four-gene motif using the FD relation (10). 
The CV is considerably lower than that in the case of the genetic toggle (figure 5). The four-gene motif, employing dual strategies of regulation, has greater noise-reduction capability 
than the genetic toggle incorporating only transcriptional regulation. 

In section 3 of the paper, we have compared transcriptional regulation (Motif 1) with post-transcriptional regulation (Motif 2) from an information theoretic perspective. 
The channel capacity {\it I}$_{opt}$ has been calculated in the small noise approximation and considering only the intrinsic component of the total noise at the output. Figures 6({\it a}) and 6({\it b}) show the plots of CV versus $\overline p$ 
for Motif 1 and Motif 2 with F$\simeq$ 0 and F$\simeq$ 1 respectively. In the first (second ) case, the CV of Motif 2 is lower (higher) than that of Motif 1. The parameter F is a measure of the fraction of microRNA molecules bound to the mRNAs. The CV of Motif 2 is found to be lower than that of Motif 1 over an extended range of F values. This 
is reflected in the plot of the channel capacity I$_{opt}$, of Motif 2, versus F in figure 7. For F $<$ F$_{c}$ = 0.8961, I$_{opt}$ is greater than I$_{opt}^{TF}$, the channel capacity of Motif 1. The case F$\simeq$ 0 corresponds to the experimental situation reported in Ref. \cite{Schmiedel}. In this limit, I$_{opt}$ = 2.91 and I$_{opt}^{TF}$ = 1.74. Instead of being limited to ON/OFF switching, transcriptional and post-transcriptional regulation could in principle designate 2$^{I_{opt}^{TF}}$ $\sim$ 2-4 and 
2$^{I_{opt}}$ $\sim$ 6-8 distinct levels, respectively, of gene expression. Because of a higher channel capacity, post-transcriptional regulation of gene expression is more efficient in information transmission than transcriptional regulation over an extended region of F values. As F approaches the value F$_{c}$, the difference between I$_{opt}$ and I$_{opt}^{TF}$ decreases but even a 
fractional bit difference in the channel capacity can give rise to non-trivial consequences in the ability of the motifs to appropriately respond to the input signal \cite{Tabba}. The computation of the channel capacity in the small noise approximation requires only a knowledge of the variance $\sigma_{y}^{2}$ of the output protein distribution. We have not considered the effect of extrinsic noise in the computation of $\sigma _{y}$. As shown in \cite{Schmiedel}, microRNA-regulated gene expression has decreased total noise at low expression but increased 
total noise at high expression as compared to the case of unregulated gene expression. The experimental results in Ref. \cite{Schmiedel} confirm the dominant contribution of intrinsic noise to the total noise over a considerable range of expression levels. 
In this range of expression levels, post-transcriptional regulation appears to be more efficient than transcriptional regulation in the transmission of information.

\section*{Appendix}
 
 \textbf{Genetic Toggle :}
 
 The differential rate equations governing the dynamics of the genetic toggle are:
 
 \begin{equation}
 \tag{A.1}
 \frac{d[M_1]}{dt}= j_{_{M_1}}-  \: \delta _{M_{1}}\:\:  [M_{1}]  + \frac{\beta_1}{{K_1}^2+{[P_2]}^2}
\end{equation}

\begin{equation}
\tag{A.2}
 \frac{d[P_1]}{dt}= j_{_{p_1}}\: [M_{1}] - \: \delta_{ P_{1}}\:\: [P_{1}] 
\end{equation}

\begin{equation}
 \tag{A.3}
 \frac{d[M_2]}{dt}= j_{_{M_2}}-  \: \delta_{ M_{2}}\:\:  [M_{2}]  + \frac{\beta_1}{{K_1}^{2}+{[P_1]}^2}
\end{equation}

\begin{equation}
\tag{A.4}
 \frac{d[P_2]}{dt}= j_{_{p_2}}\: [M_{2}] - \: \delta_{ P_{2}}\:\: [P_{2}] 
\end{equation}

The symbols {\it M}$_{i}$ ({\it i} = 1, 2), {\it P}$_{i}$ ({\it i} = 1, 2) represent the concentrations of the mRNAs and proteins respectively. The first terms 
in the equations (A.1)-(A.4) denote unregulated synthesis rates and the second terms the degradation rates of the biomolecules. The third terms in (A.1) 
and (A.3) represent the mutual repression of transcriptional initiation. 

The composite reactions considered are: 

\begin{equation}
 \tag{A.5}
 M_{1}  \:\:\:\:  \xrightarrow{j_{_{M_1}}+\frac{\beta_1}{{K_1}^{2}+{[P_2]}^2}} \:\:\:\:  M_{1} +1
\end{equation}

\begin{equation}
 \tag{A.6}
 M_{1}  \:\:\:\:  \xrightarrow{\delta _{M_{1}}\:  [M_{1}]} \:\:\:\:  M_{1} - 1
\end{equation}

\begin{equation}
 \tag{A.7}
 P_{1}  \:\:\:\:  \xrightarrow{j_{_{p_1}} \:[M_{1}]} \:\:\:\:  P_{1} +1
\end{equation}

\begin{equation}
 \tag{A.8}
 P_{1}  \:\:\:\:  \xrightarrow{\delta _{P_{1}}\:  [P_{1}]} \:\:\:\:  P_{1} - 1
\end{equation}
and similar reactions for {\it M}$_{2}$, {\it P}$_{2}$. The elements of the reaction propensity vector {\it f}  appear over the arrows. The stoichiometric matrix 
\begin{equation}
\tag{A.9}
  S =  \left[ \begin{array}{cccccccc}
1 & -1 & 0 & 0 & 0 & 0 & 0 & 0 \\
0 & 0 & 1 & -1 & 0 & 0 & 0 & 0 \\
0 & 0 & 0 & 0 & 1 & -1 & 0 & 0 \\
0 & 0 & 0 & 0 & 0 & 0 & 1 & -1 \\\end{array} \right] 
 \end{equation}
The reaction propensity vector 
\begin{equation}
\tag{A.10}
\begin{aligned}
\textbf{f} =  (\:\:j_{_{[M_1]}}+\frac{\beta_1}{{K_1}^{2}+{[P_2]}^2} \:\:\:\:\: \delta _{M_{1}}\:  [M_{1}]\:\:\:\:j_{_{p_1}} \:[M_{1}] \:\:\:\:\: \delta _{P_{1}}\:  [P_{1}]\\
j_{_{M_2}}+\frac{\beta_2}{{K_2}^{2}+{[P_1]}^2} \:\:\:\: \delta _{M_{2}}\:  [M_{2}]\:\:\:\:j_{_{p_2}} \:[M_{2}] \:\:\:\: \delta _{P_{2}}\:  [P_{2}])^{{\it T}}
\end{aligned}
 \end{equation}
 where {\it T} denotes the transpose
 
 \textbf{Four-gene Motif : }
 
 The differential rate equations governing the dynamics of the four-gene motif are given in equations \newline (1) - (8). The reaction scheme is given by
 \begin{equation}
 \tag{A.11}
 m_{1}  \:\:\:\:  \xrightarrow{\alpha _{m{_{1}}}}\:\:\:\:\:  m{_{1}} + 1
\end{equation}

\begin{equation}
 \tag{A.12}
 m_{1}  \:\:\:\:  \xrightarrow{\gamma _{m{_{1}}} [m{_{_1}}]}\:\:\:\:\:  m{_{1}} - 1
 \end{equation}
 
 \begin{equation}
 \tag{A.13}
 m_{1},\: s_{1},\:  c_{1}  \:\:\:\:  \xrightarrow{\mu{_{1}}\:[m{_{1}}][s{_{1}}]}\:\:\:\:\:  m{_{1}} - 1, \: s{_{1}} - 1,\: c{_{1}} + 1
 \end{equation}
 
 \begin{equation}
 \tag{A.14}
 m_{1},\: s_{1},\:  c_{1}  \:\:\:\:  \xrightarrow{\mu{_{2}}[c{_{1}}]}\:\:\:\:\:  m{_{1}} + 1, \: s{_{1}} + 1,\: c{_{1}} - 1
 \end{equation}

 \begin{equation}
 \tag{A.15}
 s_{1}  \:\:\:\:  \xrightarrow{\alpha _{s{_{1}}}}\:\:\:\:\:  s{_{1}} + 1
 \end{equation}
 
 \begin{equation}
 \tag{A.16}
 s_{1}  \:\:\:\:  \xrightarrow{\gamma _{s{_{1}}}[s_{1}]}\:\:\:\:\:  s{_{1}} - 1
 \end{equation}
 
 \begin{equation}
 \tag{A.17}
 s_{1},\: c_{1}  \:\:\:\:  \xrightarrow{k_{1} [c_{1}]}\:\:\:\:\:  s{_{1}} + 1,\: c{_{1}} - 1
 \end{equation}
 
 \begin{equation}
 \tag{A.18}
 s_{1}  \:\:\:\:  \xrightarrow{\frac{t_{1}[ p_{2}]^{2}}{R_{1}+[p_{2}]^{2}}}\:\:\:\:\:  s{_{1}} + 1
 \end{equation}
 
 \begin{equation}
 \tag{A.19}
 p_{1}  \:\:\:\:  \xrightarrow{\alpha _{p{_{1}}} [m{_{_1}}]}\:\:\:\:\:  p{_{1}} + 1
 \end{equation}
 
 \begin{equation}
 \tag{A.20}
 p_{1}  \:\:\:\:  \xrightarrow{\gamma _{p{_{1}}}  [p{_{_1}}]}\:\:\:\:\:  p{_{1}} - 1
 \end{equation}
 
 \begin{equation}
 \tag{A.21}
 c_{1}  \:\:\:\:  \xrightarrow{t_{a _{1}}  [c{_{_1}}]}\:\:\:\:\:  c{_{1}} - 1
 \end{equation}
 
 \begin{equation}
 \tag{A.22}
 m_{2}  \:\:\:\:  \xrightarrow{\alpha _{m{_{2}}}}\:\:\:\:\:  m{_{2}} + 1
\end{equation}

\begin{equation}
 \tag{A.23}
 m_{2}  \:\:\:\:  \xrightarrow{\gamma _{m{_{2}}}  [m{_{_2}}]}\:\:\:\:\:  m{_{2}} - 1
 \end{equation}
 
 \begin{equation}
 \tag{A.24}
 m_{2},\: s_{2},\:  c_{2}  \:\:\:\:  \xrightarrow{\mu{_{1}}[m{_{2}}]  [s{_{2}}]}\:\:\:\:\:  m{_{2}} - 1, \: s{_{2}} - 1,\: c{_{2}} + 1
 \end{equation}
 
 \begin{equation}
 \tag{A.25}
 m_{2},\: s_{2},\:  c_{2}  \:\:\:\:  \xrightarrow{\mu{_{2}}[c{_{2}}]}\:\:\:\:\:  m{_{2}} + 1, \: s{_{2}} + 1,\: c{_{2}} - 1
 \end{equation}

 \begin{equation}
 \tag{A.26}
 s_{2}  \:\:\:\:  \xrightarrow{\alpha _{s{_{2}}}}\:\:\:\:\:  s{_{2}} + 1
 \end{equation}
 
 \begin{equation}
 \tag{A.27}
 s_{2}  \:\:\:\:  \xrightarrow{\gamma _{s{_{2}}}[s_{2}]}\:\:\:\:\:  s{_{2}} - 1
 \end{equation}
 
 \begin{equation}
 \tag{A.28}
 s_{2},\: c_{2}  \:\:\:\:  \xrightarrow{k_{2} [c_{2}]}\:\:\:\:\:  s{_{2}} + 1,\: c{_{2}} - 1
 \end{equation}
 
 \begin{equation}
 \tag{A.29}
 s_{2}  \:\:\:\:  \xrightarrow{\frac{t_{2}[ p_{1}]^{2}}{R_{2}+[p_{1}]^{2}}}\:\:\:\:\:  s{_{2}} + 1
 \end{equation}
 
 \begin{equation}
 \tag{A.30}
 p_{2}  \:\:\:\:  \xrightarrow{\alpha _{p{_{2}}}  [m{_{_2}}]}\:\:\:\:\:  p{_{2}} + 1
 \end{equation}
 
 \begin{equation}
 \tag{A.31}
 p_{2}  \:\:\:\:  \xrightarrow{\gamma _{p{_{2}}}  [p{_{_2}}]}\:\:\:\:\:  p{_{2}} - 1
 \end{equation}
 
 \begin{equation}
 \tag{A.32}
 c_{2}  \:\:\:\:  \xrightarrow{t_{a _{2}} [c{_{_2}}]}\:\:\:\:\:  c{_{2}} - 1
 \end{equation}
 
 The entries over the arrows in the reaction set \newline (A.11)-(A.32) constitute the successive elements the column vector \textbf { f} ( reaction propensity vector ). 
 The stoichiometric matrix \textbf{S} is a 8$\times$22 matrix the elements of which are specified by the reaction scheme (A.11)-(A.32).
 \subsection*{Acknowledgments}

MP acknowledges the support by UGC, India, vide sanction Lett. No. F.2-8/2002(SA-I) dated 23.11.2011. IB acknowledges the support by CSIR, India, vide sanction Lett. No. 
21(0956)/13-EMR-II dated 28.04.2014.

 \end{document}